# Continuously wavelength-tunable high harmonic generation via soliton dynamics

FRANCESCO TANI,[1,*] MICHAEL H. FROSZ,[1] JOHN C. TRAVERS,[1,2] AND PHILIP ST.J. RUSSELL[1,3]

[1]*Max Planck Institute for the Science of Light, Staudtstr. 2, 91058 Erlangen, Germany*
[2]*School of Engineering and Physical Sciences, Heriot-Watt University, Edinburgh, EH14 4AS, UK*
[3]*Department of Physics, University Erlangen–Nuremberg, Erlangen, Germany*
*\* Corresponding author: francesco.tani@mpl.mpg.de*

**We report generation of high harmonics in a gas-jet pumped by pulses self-compressed in a He-filled hollow-core photonic crystal fiber through the soliton effect. The gas-jet is placed directly at the fiber output. As the energy increases the ionization-induced soliton blue-shift is transferred to the high harmonics, leading to a emission bands that are continuously tunable from 17 to 45 eV.**

http://dx.doi.org/10.1364/OL.99.099999

Soliton dynamics offer a range of powerful tools for nonlinear manipulation of ultrashort light pulses. In gas-filled hollow-core photonic crystal fiber (HC–PCF), the weak anomalous dispersion gives access to these dynamics over a broad spectral range in the visible and near infrared regions. Moreover, the pressure-tunable dispersion adds an additional degree for controlling the interplay between linear and nonlinear effects [1]. In addition to the Kerr effect, temporally non-local nonlinearities such as Raman scattering and ionization offer additional tools for manipulating guided light pulses [2–4]. By making use of these dynamics, gas–filled HC-PCFs have been successfully exploited for supercontinuum generation, efficient generation of broadband ultraviolet and vacuum ultraviolet (VUV) radiation and soliton self–compression down to sub-single cycle durations [5–7] – an ideal pump source for high harmonic generation (HHG) [8, 9].

HHG is an extreme nonlinear process, which depends on many parameters, such as the peak intensity, the duration and the global phase of the driving pulse as well as the phase matching conditions. By controlling these parameters it is possible to manipulate the spectrum of the emitted extreme ultraviolet (XUV) radiation, and perhaps even the temporal profile of the associated attosecond pulses [10]–[13]. Manipulation of high harmonic spectra has been achieved previously by exploiting various different mechanisms, such as controlling the temporal chirp of the driving laser, the ionization-induced blue shift of the driver pulse in the generating medium [2, 10, 13], and tuning the central wavelength of the driver using an optical parametric amplifier [14]. Here we combine a gas-filled HC–PCF with a gas–jet for HHG driven by laser pulses of a few tens of μJ, which undergo soliton self–compression in the fiber. We show that by exploiting the interaction of the soliton with the ionization current in the waveguide we can continuously upshift the soliton frequency [4, 15, 16] and as a result tune the frequency of the generated harmonics, achieving in this way almost 3 eV of blue-shift – larger than reported in previous work (1–2 eV) [2,10] while at the same time compressing the driving pulse. The high efficiency and low pulse energies suggest that this HHG scheme will pave the way to the development of a tunable XUV source that can easily be scaled to MHz repetition rates (soliton self–compression in gas-filled HC-PCF was previously demonstrated at 38 MHz repetition rate [17]).

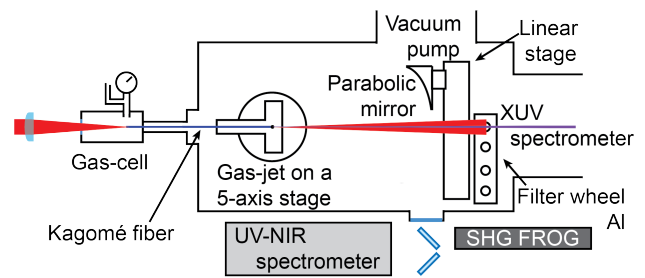

**Fig. 1.** Sketch showing the experimental arrangement.

In the experiment, pulses of duration 25 fs at 800 nm (1.5 eV) with 1 kHz repetition rate and energies in the range 10-56 μJ (Femtolasers FEMTOPOWER PRO HE CEP), were launched into a 26 cm long kagomé-style hollow-core PCF with 46 μm core diameter (Fig 1(a)), using an achromatic lens with 20 cm focal length. The pulse energy was varied using a half–wave plate and a thin-film polariser. The dispersion introduced by the optics was pre-compensated by a pair of chirped mirrors providing 250 fs$^2$ per bounce and further adjusted by tuning the laser grating compressor. One end of the fiber was enclosed in a gas cell filled with 5 bar of He, while the other was fed into a vacuum chamber, which—because of the small fiber core—could be evacuated down to 10$^{-5}$ mbar using a turbo-molecular pump with a pumping speed of 250 l/s [18].

At the output, the fiber was precisely positioned perpendicular to a pulsed gas-jet emitted from a nozzle 200 µm in diameter. Two 200 nm thick aluminium filters were placed on the other side of the jet, aligned so as to face the fiber endface, and the spectrum of the isolated HHG beam was measured using a flat-field spectrometer (Dr Hörlein & partner). The pulsed gas-jet was placed on a 5-axis stage, making it possible to adjust the position of the nozzle with respect to the fiber endface. It was supplied with Ar gas at 5 bar using a piezo-valve that was synchronised with the laser pulses and the CCD camera of the XUV spectrometer. The distance between the fiber axis and the nozzle was ~270 µm and the pressure in the jet was estimated (following [19]) to be ~100 mbar. When the gas-jet was operating, the pressure in the vacuum chamber rose up to a few times 10$^{-3}$ mbar, resulting in ~20% absorption of the XUV radiation before the detector, which was 1.2 m from the gas-jet. A mirror was inserted after the fiber and before the aluminium filters so as to deliver the beam, through a 3.3 mm CaF$_2$ viewport outside the vacuum chamber, for diagnostic tests of the fundamental laser pulse after propagation through the fiber. This mirror was also used to maximise the coupling of the pump beam into the fundamental mode of the fiber, achieving in this way ~84% transmission which, considering the 2 dB/m fiber loss at 800 nm, translates to ~94% launch efficiency. We could launch over 70 µJ into the core without damage. When damage did occur at higher energies, this was mainly caused by beam-pointing stabilities in the laser. Note that the launch efficiency did not degrade when we filled the gas-cell with 5 bar He, since the nonlinearity of this gas is very low and the ionization potential very high.

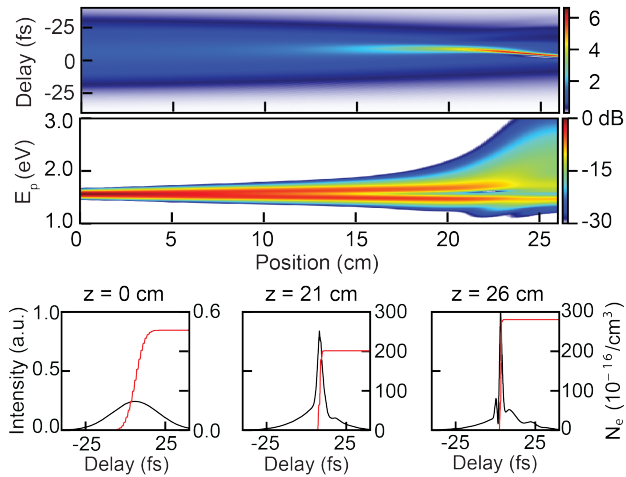

**Fig. 2.** Temporal and spectral evolution of a 25 fs pulse with 50 µJ energy and positively chirped by 210 fs$^2$ undergoing soliton self-compression in HC–PCF filled at the input with 5 bar He. Snapshots of the pulse and the corresponding free electron densities (simulation).

The pulse dynamics in the fiber depend strongly on the He filling pressure, which was selected so that the pulses were propagating in the anomalous dispersion regime with soliton order $N = \sqrt{\gamma P_0 T_0^2/|\beta_2|}$ between 2 and ~5 (depending on the launched pump energy), where $\gamma$ is the nonlinear coefficient, $P_0$ the soliton peak power, $T_0$ the soliton duration and $\beta_2$ the group velocity dispersion. Under these conditions, for constant pressure along the fiber, the pump pulse experiences clean adiabatic soliton self-compression as it propagates, with a compression factor inversely proportional to $N$ [1]. The length over which the input laser pulse compresses is proportional to $L_d/N$, where $L_d = T_0^2/\beta_2$ is the dispersion length. In the presence of a pressure gradient self-compression occurs over a longer distance. In both cases, the compression length can be adjusted to match precisely the fiber length, without strongly affecting the final pulse duration, by adding a positive chirp to the input pulse. In the experiment we match the two lengths by positively chirping the pulses launched in the fiber by ~200 fs$^2$. As the energy increases the pulse gets shorter, its peak intensity rises and the influence of ionization becomes more important, eventually causing the self-compressing pulse to blue-shift in frequency. The magnitude of this shift is given by [10, 20]:

$$\delta\omega = \frac{e^2 L}{2cn_0 m_e \varepsilon_0 \omega} \frac{\partial N_e}{\partial t},$$

where $c$ is the speed of light, $m_e$ the electron mass, $\varepsilon_0$ the vacuum permittivity, $e$ is the electronic charge and $L$ is the size of the plasma. The equation shows that the shift is proportional to $L$ and to the slope of the free electron density $\partial N_e/\partial t$, so is larger for shorter pulses. As a result of the long interaction length over which a short pulse duration is sustained, a very large frequency shift can be achieved through self-compression. The dynamics are described by the plots in Fig. 2, which show the temporal and spectral evolution of a 25 fs pulse with 50 µJ energy and a 210 fs$^2$ positive chirp as it propagates along the fiber. The simulations were based on a unidirectional field equation [21] including photoionization with the ADK model [22]. They predict that for these parameters the pulse should compress down to sub-3 fs. The lower plots show snapshots of the pulse together with the relative free electron density. As the pulse compresses, the free electron density increases and when it becomes high enough (at ~20 cm), the pulse accelerates and blue-shifts.

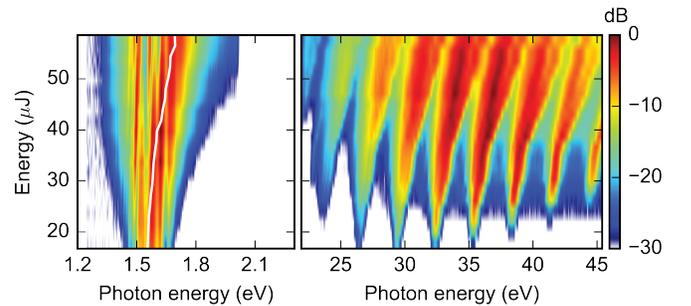

**Fig. 3**. Measured pump pulse (left) and high-harmonic (right) spectra for increasing launched pulse energy. The white curve on the left-hand side tracks the frequency of the 17$^{th}$ harmonic divided by 17.

The resulting compressed pulses have sufficient intensity to generate high harmonics in the gas–jet and, because the jet is placed after the fiber endface, short trajectories are favored. Fig. 3 plots the pulse spectra at the output of the fiber (left) together with the high harmonic spectra (right) for increasing pump pulse energy. The driver pulse spectra were measured with the gas–jet switched off, confirming that the pulse blue-shifts as it propagates along the gas-filled fiber. The harmonic spectra were integrated along the spatial axis of the CCD. The bandwidth of each harmonic grows with increasing pulse energy, suggesting temporal compression of the driving pulses. Along with spectral broadening, there is a clear correlation between the pump pulse spectra and the high harmonic spectra: the harmonic frequencies track the blue-shifting soliton driver as the input energy is increased.

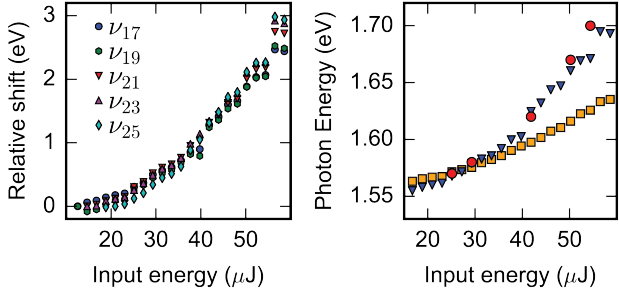

**Figure 4**. Frequency shifts of the 17th – 25th harmonics relative to the frequency at the lowest input energy, plotted as a function of pump energy (left). The plot on the right panel shows the central frequency of the pump pulse (squares), the averaged frequency of the 17th harmonic divided by 17 (triangles) and the soliton central frequency (circles) obtained from the measured spectrograms.

The harmonics do not follow the average wavelength of the pulses launched into the PCF ($\nu_{pump}$), but rather the central wavelength of the blue-shifting soliton ($\nu_{sol}$). This indicates that the generation of the high-order harmonics is predominantly driven by the self-compressing blue-shifting pulse, which contains most of the energy. To highlight this, a white curve, representing the average frequency of the 17th harmonic divided by its harmonic order ($\nu_{17}/17$), is drawn on top of the pump pulse spectra (Fig. 3(left)). This curve clearly follows the frequency of the blue-shifting soliton. Further support is evident in the right-hand plot of Fig. 4, where $\nu_{17}/17$ clearly does not follow $\nu_{pump}$, but rather $\nu_{sol}$ (red dots), which were obtained from spectrograms measured at different energies (see below). The shift to higher frequency is the same for all the harmonics, as shown in the left-hand plot in 4, and amounts to more than 100% of the adjacent harmonic spacing, permitting the harmonic spectra to be tuned continuously from at least 17 eV to 45 eV.

For pump energies below ~25 μJ, a few harmonics are generated and only a weak spectral broadening can be observed. As the energy in the fiber is increased up to ~40 μJ, the harmonics are generated more efficiently and the spectra start to broaden asymmetrically to the high frequency side. However, at these energies, the compression length exceeds the fiber length and highly chirped pulses are delivered to the gas–jet, resulting in a frequency shift of the harmonics, in agreement with the observations reported in [12]. For higher energies, the two lengths become comparable and the self-compressing pulses blue-shift within the fiber because of the gas–ionization, resulting in a much more pronounced frequency shift and spectral broadening of the harmonics. The right-hand plot in Fig. 4 shows the central wavelength of the pump pulse for increasing energy, the corresponding frequency shift of the 17th harmonic (which becomes much more marked for energies above 40 μJ), and the soliton central frequency, which agrees perfectly with the harmonic frequency shift.

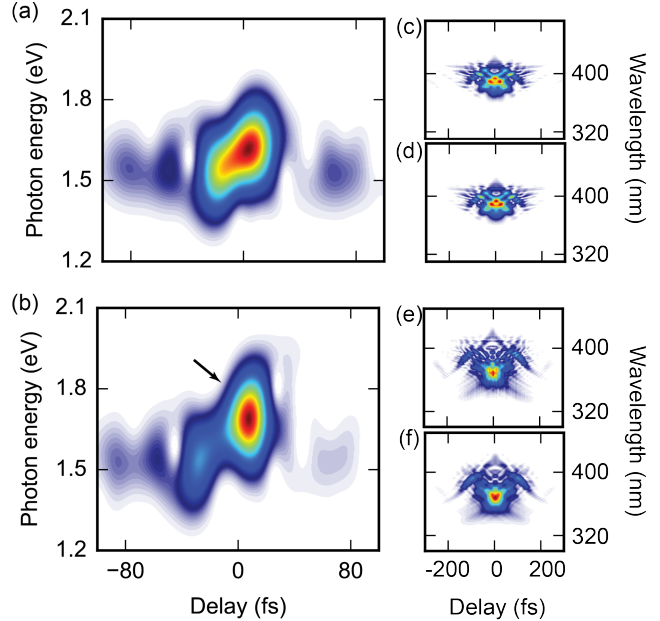

**Figure 5**. Reconstructed spectrogram for (a) 29 μJ and (b) 50 μJ launched pulse energy. (c) Measured and (d) retrieved FROG traces at the fiber output for 29 μJ pulse energy. (e) & (f) The same for 50 μJ pulse energy. A linear scale is used for all the plots. The black arrow indicates the blue shifting pulse.

These dynamics are further confirmed by the reconstructed spectrograms shown in Fig. 5, which were obtained from frequency resolved optical gating (FROG) traces measured using a home-built second harmonic (SH) FROG with a 10 μm thick BBO crystal. The FROG was placed outside the vacuum chamber, and the traces in Fig. 5 correspond to launching pulses with 29 μJ and 50 μJ energy into the fiber. In the reconstruction of the spectrograms we compensated for the dispersion introduced by the vacuum chamber window (3.3 mm $CaF_2$), the air path (1.4 m) and the half-waveplate (840 μm $SiO_2$ + 675 μm $CaF_2$), and used a 10 fs (FWHM) Gaussian pulse as the spectrogram gate function. After compensation, we obtained respectively ~38 fs and ~15 fs. The pulse compression is mainly limited by a band of fiber loss at ~2 eV (~625 nm), which also causes the sharp cut-off in the pump spectrum for energies above ~45 μJ. Note that the time direction ambiguity of SH FROG is removed by the knowledge of the dispersive elements placed before the temporal characterisation.

In agreement with the discussion above, when a 29 μJ pulse is launched in the fiber, the output spectrogram (Fig. 5 (a)) reveals a strongly chirped pulse, while in the 50 μJ case, the spectrogram clearly shows that most of the energy is carried by a blue-shifting pulse with a minor chirp, only a small fraction remaining at ~1.51 eV (~810 nm). This confirms that the high harmonics must be predominantly generated by the blue-shifting pulse.

In conclusion, high harmonics can be generated in an Ar gas jet using pump pulses self–compressed by soliton effects in gas–filled HC–PCF. In previous work [2, 10, 11, 12], frequency tuning of the harmonics was achieved by exploiting the ionization of the generating medium or by chirping the pump pulse. In both cases, the two processes: the generation and the tuning of the harmonics were strongly coupled with each other. In contrast, the soliton dynamics in the presence of ionization in the gas-filled fiber permits frequency up-shift of the pump pulse before HHG resulting in an independent continuous tuning of the high harmonics over a frequency range greater than the spacing between neighboring harmonic orders.